\documentclass[twocolumn,showpacs,preprintnumbers,amsmath,amssymb,superscriptaddress]{revtex4}

\usepackage{epsf}
\usepackage{graphicx}
\usepackage{sidecap}

\usepackage{color}

\def \beq {\begin{equation}}
\def \eeq {\end{equation}}
\begin{document}

\title{Dirac State in a Centrosymmetric Superconductor $\alpha$-PdBi$_2$}

\author{Klauss~Dimitri}\affiliation {Department of Physics, University of Central Florida, Orlando, Florida 32816, USA}

\author{M.~Mofazzel~Hosen}\affiliation {Department of Physics, University of Central Florida, Orlando, Florida 32816, USA}

\author{Gyanendra~Dhakal}\affiliation {Department of Physics, University of Central Florida, Orlando, Florida 32816, USA}

 \author{Hongchul~Choi} \affiliation{Theoretical Division, Los Alamos National Laboratory, Los Alamos, New Mexico 87545, USA}

\author{Firoza~Kabir}\affiliation {Department of Physics, University of Central Florida, Orlando, Florida 32816, USA}

\author{Dariusz~Kaczorowski} \affiliation{Institute of Low Temperature and Structure Research,
Polish Academy of Sciences, 50-950 Wroc\l aw, Poland}

\author{Tomasz Durakiewicz} \affiliation{Condensed Matter and Magnet Science Group, Los Alamos National Laboratory, Los Alamos, NM 87545, USA}

\affiliation {Institute of Physics, Maria Curie - Sk\l odowska University, 20-031 Lublin, Poland}
\author{Jian-Xin~Zhu} \affiliation{Theoretical Division, Los Alamos National Laboratory, Los Alamos, New Mexico 87545, USA}

\affiliation{Center for Integrated Nanotechnologies, Los Alamos National Laboratory, Los Alamos, New Mexico 87545, USA}

\author{Madhab~Neupane}
\affiliation {Department of Physics, University of Central Florida, Orlando, Florida 32816, USA}

\date{\today}
\pacs{}

\begin{abstract}

{
Topological superconductor (TSC) hosting Majorana fermions has been established as a milestone that may shift our scientific trajectory from research to applications in topological quantum computing. Recently, superconducting Pd-Bi binaries have attracted great attention as a possible medium for the TSC phase as a result of their large spin-orbit coupling strength. Here, we report a systematic high-resolution angle-resolved photoemission spectroscopy (ARPES) study on the normal state electronic structure of superconducting $\alpha$-PdBi$_2$ ($T_{c}$ = 1.7 K). Our results show the presence of Dirac states at higher-binding energy with the location of the Dirac point at 1.26 eV below the chemical potential at the zone center. Furthermore, the ARPES data indicate multiple band crossings at the chemical potential, consistent with the metallic behavior of $\alpha$-PdBi$_2$. Our detailed experimental studies are complemented by first-principles calculations, which reveal the presence of surface Rashba states residing in the vicinity of the chemical potential. The obtained results provide an opportunity to investigate the relationship between superconductivity and topology, as well as explore pathways to possible future platforms for topological quantum computing.
}
\end{abstract}
\maketitle

Topological insulators (TIs) are a class of quantum materials, which behave as insulators in the bulk, yet possess gapless spin-polarized surface states, which are robust against non-magnetic impurities \cite{Hasan, SCZhang}. The unique properties of TIs make them attractive not only for studying various fundamental phenomena in condensed matter and particle physics, but also as promising candidates for applications ranging from spintronics to quantum computation \cite{SCZhang, Neupane3, Wang3, Neupane4, Yang, Neupane5, Young, Hasan, Mong, Tang}. Specifically, it has been recently predicted that long-sought-out Majorana fermions can be realized at the interface between a topological insulator and a superconductor \cite{Fu, Qi, Sato, Vorontsov, Tanaka, Sato2}. These intriguing quasiparticles may be utilized for realization of robust qubits, necessary to develop quantum devices while avoiding decoherence, which limits the lifetimes of useful states in currently available, non-topological, qubits. The prospective for potential discoveries has ignited much interest in material candidates that may host Majorana fermions \cite{Kitaev2, He}. Recently, there has been a great effort in condensed matter physics to realize the Majorana fermion quasiparticle states \cite{Delft, Ali}. 
Though the experimental realizations of topological superconductivity (TSC) states in real materials remain considerably limited, the search for bulk topological superconductivity has already encompassed several classes of materials, such as Cu- or Sr-intercalated Bi$_2$Se$_3$ \cite{Hor, Kriener, Wray, Sasaki, Liu, Shruti, Neupane}, In-doped SnTe \cite{Sasaki2}, highly-pressured Bi$_2$Te$_3$ and Sb$_2$Te$_3$ \cite{Zhang, Zhu2}, point-contact regions in Cd$_3$As$_2$ \cite{Wang}, hybrid superconductor--semiconductor and superconductor--topological insulator heterostructures \cite{Delft, Su}, as well as some half-Heusler compounds \cite{Chadov, Lin}.

\begin{figure*}
\centering
\includegraphics[width=14 cm]{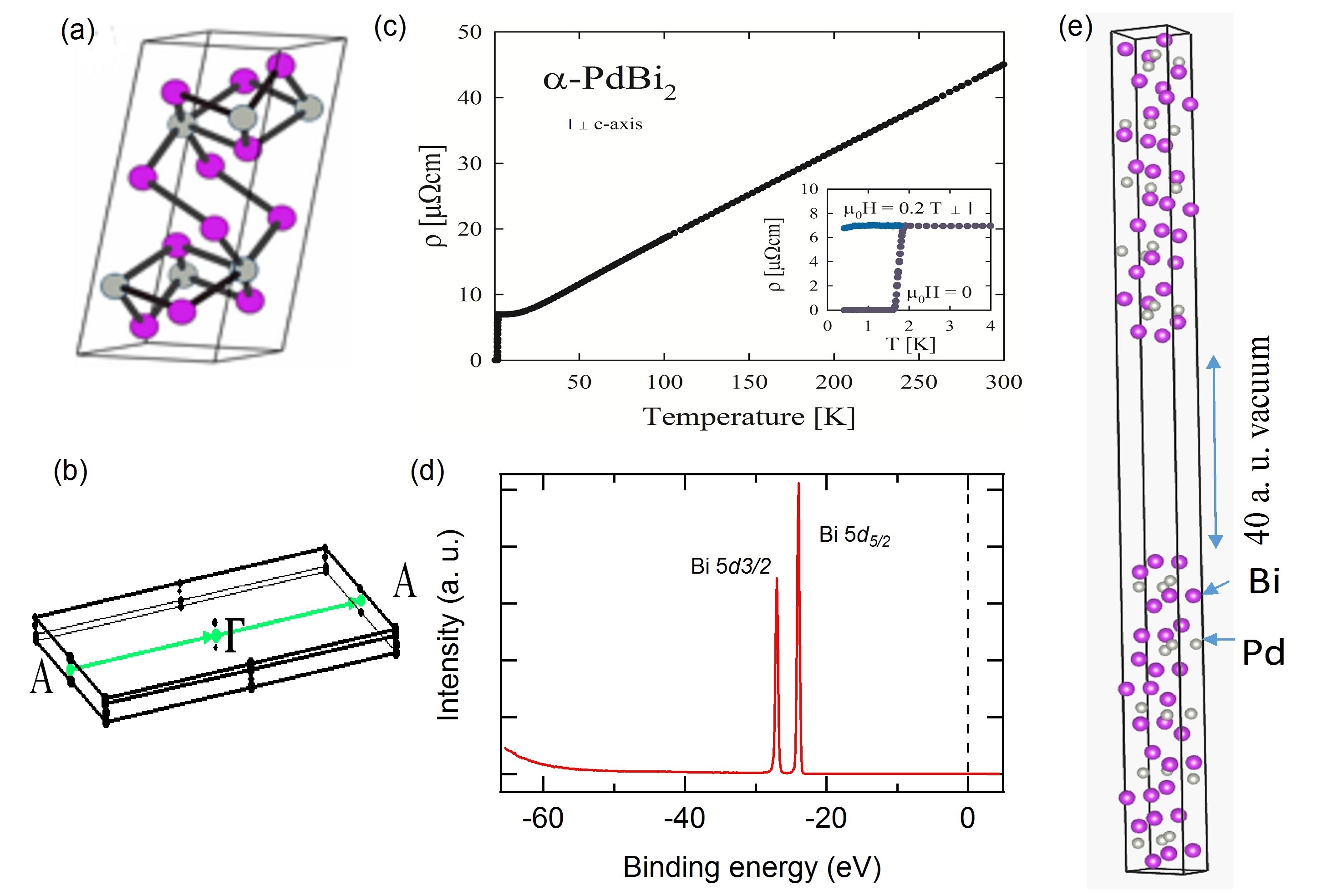}
\caption{Crystal structure, electrical transport and spectroscopic characterization of  $\alpha$-PdBi$_2$.
(a) Crystal structure of $\alpha$-PdBi$_2$; purple and gray balls represent Bi and Pd atoms, respectively. (b) 2D Brillouin zone and momentum path used for the band-structure calculation in the $5 \times1 \times 1 $ supercell of $\alpha$-PdBi$_2$. (c) Temperature dependent resistivity of $\alpha$-PdBi$_2$. Inset shows the low-temperature resistivity data taken in zero magnetic field and in a transverse magnetic field of 0.2 T. (d) Core level spectrum of $\alpha$-PdBi$_2$. (e) The $5 \times 1 \times 1$ supercell of $\alpha$-PdBi$_2$; purple (gray) balls represent Bi(Pd) atoms.}
\end{figure*}

The Pd-Bi family of compounds have been identified as potential materials to explore topological superconductivity due to their intrinsic capability to maintain strong spin orbit coupling (SOC). The noncentrosymmetric $\alpha$-PdBi with superconducting transition at $T_c$ = 3.8 K shows Dirac cone electronic dispersion at 700 meV below the chemical potential, which negates the possible TSC on the surface of this material \cite{ Sun,  Neupane2, S,  Wahl}. If the superconducting topological insulating phase is realized in $\alpha$-PdBi, one can expect to identify a mixture of singlet and triplet pairing due to the noncentrosymmetic structure and strong SOC \cite{Joshi, Neupane2}. The compound $\beta$-PdBi$_2$ is another member of the Pd-Bi family that has been reported to possibly host the TSC below $T_c$ = 5.3 K; however, the Dirac point in this material is located at 2.41 eV below the chemical potential \cite{Sakano} and lack of TSC state has been indicated by point-contact Andreev refection spectroscopy \cite{Che}. The surface states found in these materials have no effect on the bulk of the superconductor \cite{Biswas}. Another member of this family $\alpha$-PdBi$_2$ with the same molecular composition as $\beta$-PdBi$_2$ has recently been suggested to possess possible surface states from penetration depth measurements \cite{Mitra} and the first-principle calculations \cite{Choi}. However, a detailed electronic structure study of $\alpha$-PdBi$_2$ using energy-momentum resolved spectroscopic technique is still lacking.

Here, we report our experimental study of the electronic structure of the $\alpha$-PdBi$_2$ single crystal in its normal state using angle-resolved photoemission spectroscopy (ARPES). We observe a Dirac state at a high binding energy with the Dirac node at around 1.26 eV below the chemical potential. In addition, we see surface states crossing the chemical potential, in line with the metallic transport characteristics of the compound, which exhibit Rashba-type splitting. These experimental results are well corroborated by first-principles calculations. Our findings negate the possibility of a TSC state at the surface of $\alpha$-PdBi$_2$ in its native state; however, Dirac states can be realized in this material by tuning the chemical potential.

\begin{figure*}
\centering
\includegraphics[width=18.3cm]{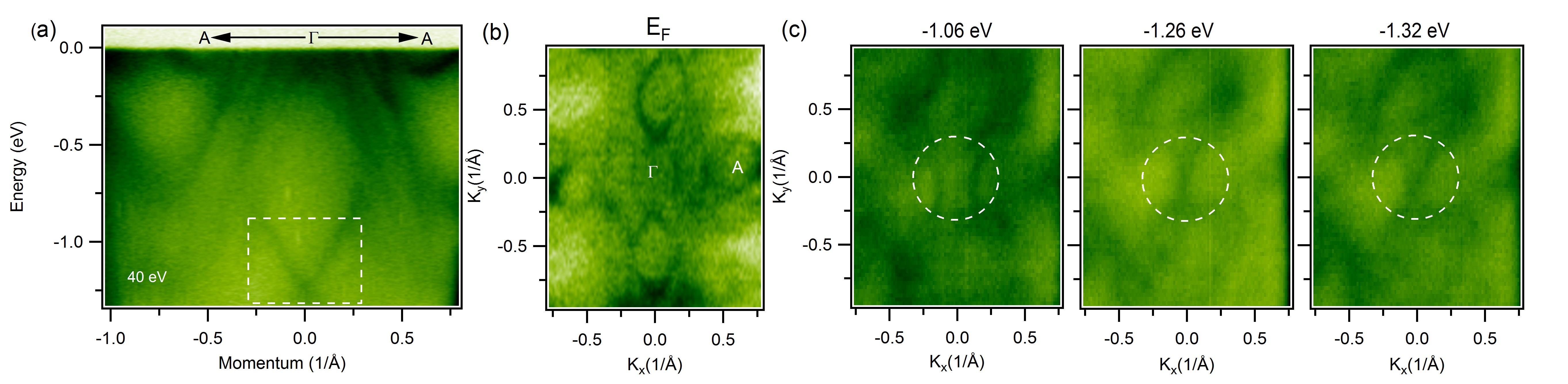}
\caption{Measured electronic structure of $\alpha$-PdBi$_2$.
(a) Dispersion map along the high symmetry A-$\Gamma$-A direction. Dashed rectangle encompasses the region of the Dirac-like point located at 1.26 eV below the chemical potential. (b) Fermi surface map determined with an incident photon energy of 50 eV at a temperature of 16 K. (c) Constant energy contour plots determined at binding energies above, near and below the Dirac point (panels from left to right). The binding energies are noted in the plots. White dashed circles encompass the Dirac states. Additional data are presented in Supplementary Materials.}
\end{figure*}

Single crystals of $\alpha$-PdBi$_2$ were obtained using a growth technique described elsewhere \cite{Mitra}. Their chemical composition and crystal structure were examined by energy-dispersive X-ray spectroscopy and single-crystal X-ray diffraction (XRD), respectively. Synchrotron-based ARPES measurements were performed at the Advanced Light Source (ALS), Berkeley at Beamline 10.0.1 and 7.0.2 equipped with a high efficiency R4000 electron analyzer. The energy resolution was better than 20 meV and the angular resolution was better than 0.2$^{\circ}$. Samples were cleaved in situ and measured between 10 - 80 K in vacuum better than 10$^{-10}$ torr. The cleavage plane of single-crystalline $\alpha$-PdBi$_2$ was (100) plane. These crystals were found to be very stable without surface degradation for the typical measurement period of 20 hours.
In order to reveal the nature of the states observed in $\alpha$-PdBi$_2$, the ARPES data were compared with the calculated band dispersions projected onto a 2D Brillouin zone (BZ). We utilized first-principles calculations of the bulk band structure as well as slab calculations performed by employing a density functional theory (DFT) with the full-potential linearized augmented plane wave (FP-LAPW) method implemented in a WIEN2k package \cite{Blaha}.

The XRD results indicated the $\alpha$-PdBi$_2$ phase with a monoclinic centrosymmetric unit cell (space group $C2/m$) and lattice parameters close to those reported for this compound in the literature: $a$ = 12.74 $\AA$, $b$ = 4.25 $\AA$, $c$ = 5.665 $\AA$, $\alpha$ = $\gamma$ = 90$^{\circ}$, and $\beta$ = 102.58$^{\circ}$ \cite{ Zhuravlev}. The crystallographic unit cell of $\alpha$-PdBi$_2$ with the $a$-axis being its unique direction is shown in Fig. 1a. The bulk Brillouin zone with symmetry at the zone center $\Gamma$ and at the edge A on the (100) surface can be seen in Fig. 1b. Quality of the obtained single crystals of $\alpha$-PdBi$_2$ was checked by measuring the electrical resistivity with electric current flowing within the (100) plane. The compound was found to exhibit good metallic conductivity with a sharp transition to the superconducting state at $T_{c}$ = 1.7 K (see Fig. 1c), as reported in Ref.~\cite{Mitra}. 
Worth noting is the residual resistivity of 7 $\mu \Omega$cm, measured just above $T_{c}$, which is distinctly smaller than that found in a recent study \cite{Mitra}. This finding indicates high quality of the crystals used in our ARPES experiments. Fig. 1d presents the momentum-integrated ARPES spectrum of $\alpha$-PdBi$_2$ recorded over a wide energy window. The two intensity peaks at binding energies $E_B$ = 24 and 27 eV correspond to the bismuth 5\textit{d}$_5$$_/$$_2$ and 5\textit{d}$_3$$_/$$_2$ energy levels. These sharp features corroborate the high quality of the crystal used in our measurement. 

Fig. 2a depicts an ARPES dispersion map in a 1.3 eV binding energy window. One finds in this figure several valence bands crossing the chemical potential. These metallic bands do not show Dirac like dispersions, however Rashba splitting can be seen in the vicinity of the chemical potential. Interestingly, a linearly dispersive Dirac cone state is observed at the Brillouin zone center point $\Gamma$ at a high binding energy (illustrated in Fig. 2a by the dashed rectangle). The linearly dispersive Dirac bands are relatively isolated from other bulk bands surrounding it. Such a characteristic is beneficial when realizing TSC as it reduces the bulk band interferences. This Dirac node is located at a binding energy of 1.26 eV below the chemical potential. Fig. 2b displays the Fermi surface map measured with a photon energy of 50 eV, where multiple elliptically shaped metallic pockets can be seen around the high symmetry points A and $\Gamma$ that represent the edges and the center of the BZ, respectively. Fig. 2c shows the constant energy contour plots measured at various binding energies. At the 1.26 eV constant energy contour plot, corresponding to the energy of the Dirac point, a pointlike feature can be noticed, however its occurrence is somewhat obscured because of multiple bulk bands present in its vicinity. Moving towards higher and lower binding energies, the constant energy contours expand (note the white dashed circles in the panels of Fig. 2c). Supplementary material provides additional data on the Fermi surface of $\alpha$-PdBi$_2$ measured at photon energies of 70 and 90 eV with highly detailed mapping where high symmetry points are clearly recognizable. 

In order to determine the nature of the linearly dispersive bands associated with the Dirac point, several energy-momentum dispersion cuts were made with different photon energies between 50 and 70 eV (see Fig. 3a) and additionally between 70 and 90 eV (see Supplementary Fig. 3). Here, we observe several metallic bands crossing the chemical potential. Importantly, the Dirac-like state is seen in all the plots, and this feature does not disperse with the photon energy, while the other metallic bands disperse with a change in the perpendicular momentum. This implies that the linearly dispersive states in $\alpha$-PdBi$_2$ originate from the surface. Similarly, the observed Rashba-type splitting is unaffected by the change in the photon energy, which asserts that the observed bands originate from the surface. Fig. 3b presents the result of our first-principles calculations made for the $ 5 \times 1 \times 1 $ supercell of $\alpha$-PdBi$_2$ (see Fig. 1e). As can be inferred from this figure, along the A-$\Gamma$-A direction in BZ there occurs a Dirac-like dispersion at the $\Gamma$ point and a Rashba splitting around the A point in the vicinity of the chemical potential. This finding is in very good agreement with the experimental data (a small energy gap seen in Fig. 3b can be attributed to size effect in the slab calculations).

\begin{figure*}
\centering
\includegraphics[width=18cm]{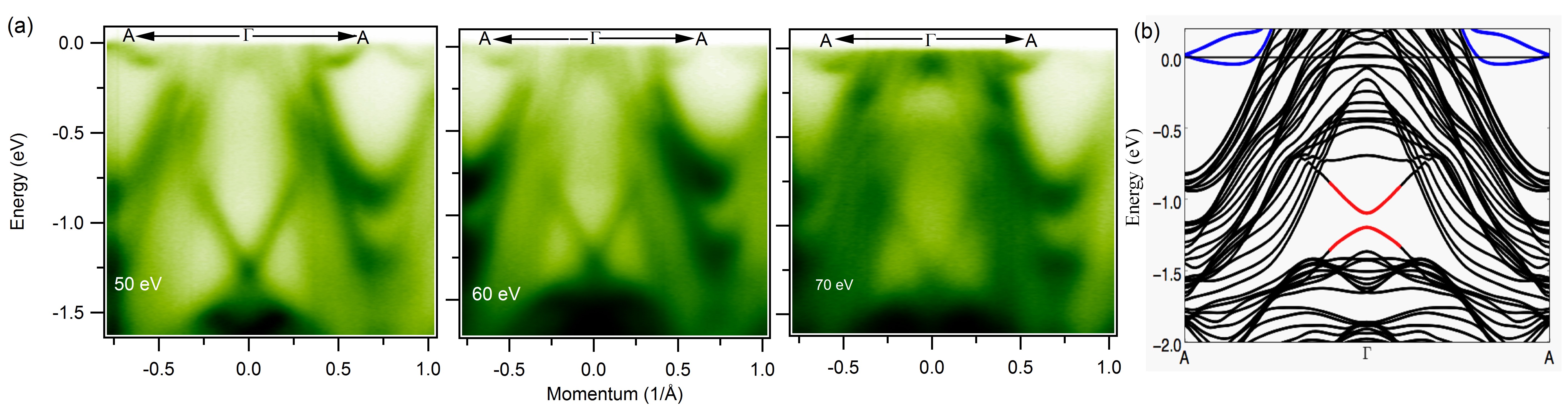}
\caption{Dirac-like state in $\alpha$-PdBi$_2$. (a) Dispersion maps along the high symmetry A-$\Gamma$-A direction obtained using photon energies of 50, 60 and 70 eV. Additional supporting data from 70 eV to 90 eV taking 2 eV steps can be found in supplementary figure 3. (b) Band structure calculated for the $ 5 \times 1 \times 1 $ supercell. The red and blue lines represent the linearly dispersive and Rashba spin-orbit bands, respectively.}
\end{figure*}

\begin{SCfigure*}
\centering
\includegraphics[width= 14.5cm]{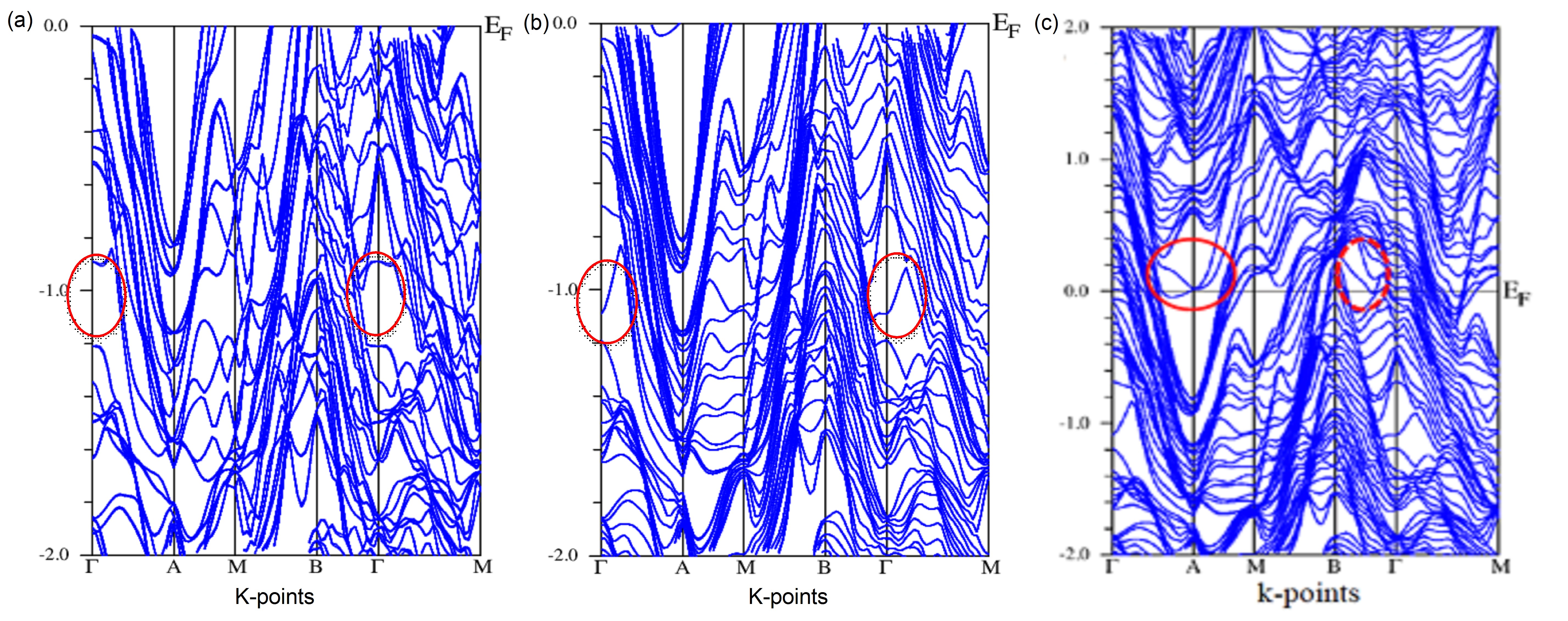}
\caption{Calculated electronic band structure of $\alpha$-PdBi$_2$.
(a) Band structure calculated without SOC. (b,c) Band structure calculated with SOC taken into account. The oval curves are provided to emphasize the effect of the SOC.}
\end{SCfigure*}

Figure 4 illustrates the electronic band structure of $\alpha$-PdBi$_2$ calculated from first principles with and without considering the spin-orbit coupling (SOC). The inclusion of SOC vividly reveals the Rashba-type splitting at the A point in the vicinity of chemical potential, as shown in Fig. 4c (a more detailed discussion of these results can be found in Ref.~\cite{Choi}). The linear dispersion bands around 1.26 eV below the Fermi level at the $\Gamma$ point are affected by SOC and the surface potential. As emphasized by the oval curves in Fig. 4a and 4b, the flat bands around 1.0 eV below the chemical potential at the $\Gamma$ point observed at the absence of SOC change into linear dispersion upon taking into account the SOC effect. It is noteworthy that the surface state is essential to make this change \cite{Choi}. Since the energy gap between the two bands gets suppressed by Rashba-type SOC, this gap may close with a larger slab geometry.

Importantly, the similarity between the measured data and the calculation results occurs at the $\Gamma$ point where the Dirac point is located at 1.26 eV below the chemical potential. As a result of observing the same band structure at this high symmetry point we are able to determine that the Dirac state originates from the surface but is affected by the surface potential. We further confirm the origin of Dirac-like dispersion with our calculations from a $5 \times 1 \times 1$ $\alpha$-PdBi$_2$ supercell. Another characteristic feature realized by both measurement and calculation is the bands structure found around the high symmetry point A. Our experimental data reveal the presence of the Rashba state near the chemical potential. The momentum-dependent splitting is induced in the degenerate surface state as a direct result of SOC driving the troughs around the A point \cite{Choi}.

In conclusion, with a systematic analysis of the electronic structure of $\alpha$-PdBi$_2$, we have identified the existence of a Dirac state that mostly originates from the surface. We find that the Dirac state is about 1.26 eV below the chemical potential. The Dirac state also exhibits ideal characteristics including its relative isolation from other bands surrounding it, which greatly reduces the issue of interference when attempting to realize a TSC state. Tuning this Dirac state towards the Fermi energy would provide the possibility of realizing TSC. We also report in $\alpha$-PdBi$_2$ some low-energy surface states that develop from Rashba splitting in close proximity to the Fermi energy, and are likely governed by the surface potential. These states eventually meet at the high symmetry point A to form band crossings. Since these states are located close to the chemical potential, slight tuning may realize spin-split Rashba states in this material. Our study of $\alpha$-PdBi$_2$ provides a great opportunity to reveal multiple phases in a superconducting material.
\\

M.N. is supported by the start-up fund from University of Central Florida. This work at LANL was supported by the U.S. DOE Contract No. DE-AC52-06NA25396 through the LANL LDRD Program (H.C. \& J.-X.Z.). The work was supported in part by the Center for Integrated Nanotechnologies, a DOE BES user facility, in partnership with the LANL Institutional Computing Program for computational resources. T.D. is supported by NSF IR/D program. D.K. was supported by the National Science Centre (Poland) under research grant 2015/18/A/ST3/00057. We thank S. K. Mo, Eli Rotenberg, Aaron Bostwick and Chris Jozwiak for beamline assistance at the, ALS, LBNL.

Correspondence and requests for materials should be addressed to M.N. (Email: Madhab.Neupane@ucf.edu)

\end{document}